\documentclass{aa}
\usepackage{graphicx}
\usepackage{txfonts}

\begin{document}

\title{\object{G337.2+0.1}: a new X-ray supernova remnant?}

\subtitle{}

\author{J.~A. Combi\inst{1,2}, P. Benaglia\inst{2,3}, G.E. Romero\inst{2,3},
\& M. Sugizaki\inst{4}
}
\institute{Departamento de F\'{\i}sica (EPS), Universidad de Ja\'en,
Campus Las Lagunillas s/n, 23071 Ja\'en, Spain
\and Instituto Argentino de Radioastronom\'{\i}a, C.C.5, (1894) Villa Elisa, Buenos Aires, Argentina
\and Facultad de Ciencias Astron\'omicas y Geof\'{\i}sicas UNLP, Paseo del Bosque, B1900FWA La Plata, Argentina
\and Stanford Linear Accelerator Center, 2575 Sand Hill Road, Menlo Park, CA 94025, USA
}

\authorrunning{Combi et~al.}

\offprints{J.~A. Combi,\\ \email{jcombi@ujaen.es}}

\date{Received / Accepted}

\abstract{ We present evidence supporting a SNR origin for the
radio source \object{G337.2+0.1}, which was discovered along the
line of sight to the Norma spiral arm in the MOST 843-MHz radio
survey. The radio source is spatially superposed to the
unidentified {\it ASCA} source \object{AX~J1635.9-4719}. An
analysis of this latter source reveals that its X-ray spectrum,
extended nature, and non-variable flux are consistent with what is
expected for a SNR. In addition, we have used HI-line observations
of the region to look for any effect of the presumed remnant on
the ISM. We have found a well-defined  minimum centered at the
position of the radio source in the velocity range of $\sim -25$
to $-19$ km s$^{-1}$. This feature appears as a sharp absorption
dip in the spectrum that might be produced when the continuum
emission from the SNR candidate is absorbed by foreground gas.
Hence we have used it to constrain the distance to the source,
which seems to be a young (age $\sim$ a few $10^3$ yr) and distant
($d\sim14$ kpc) SNR.
\object{G337.2+0.1} and
\object{AX~J1635.9-4719} would be the radio/X-ray manifestations
of this remnant.

 } \maketitle

\section{Introduction} \label{introduction}

Supernova remnants (SNRs) have long been considered as a primary
source of galactic cosmic rays (CRs) with energies up to the knee
of the spectrum at $\sim$ 10$^{15.5}$ eV (e.g. Shklovskii
\cite{Shkl53}, Ginzburg \& Syrovatskii \cite{Ginz64}). First-order
Fermi shock acceleration has been suggested as the most likely
acceleration mechanism for charged particles in the shells of SNRs
(e.g. Bell \cite{Bell78}, Reynolds \& Chevalier \cite{reynold81}).
Evidence supporting the presence of TeV electrons in these objects
comes from the detection of synchrotron X-rays in a number of
sources such as G347.3$-$0.5 (Koyama et al. ~\cite{koyama97}), SN 1006 (\cite{koya95}), Cas A
(Allen et al. ~\cite{allen97}), and RC W86 (Borkowsky et al.~\cite{bork01}).

A decade ago the detection of SNRs at X-rays was difficult mainly
due to the absorption of the soft (i.e., $<$ 3 keV) X-ray emission
by the large column density of gas and dust in the galactic plane.
In recent years, with the advent of the new generation of X-ray
satellites, the number of galactic SNRs detected in the energy
range above 3 keV has been increased considerably. Although the
most complete catalog of SNRs is still compiled in the radio band
(Green \cite{green04}), X-ray instruments such as {\it ASCA}, {\it
XMM} and {\it CHANDRA} have provided an important new window to
find yet undetected remnants or to confirm candidates originally
found at other wavelengths.

In this paper we report the results of a multiwavelength study of
the  field containing the unidentified X-ray source
\object{AX~J1635.9-4719} and the SNR candidate
\object{G337.2+0.1}. In order to investigate the nature and
possible physical connection between both sources, we have
performed source cross-identifications with all available
astronomical databases and we have re-processed all the relevant
data. Our results support the identification of
\object{G337.2+0.1} as a new X-ray emitting SNR. In the next
section we present a re-analysis of the {\it ASCA} data that
confirms the extended nature of the source and its non-variable
nature.  In Section ~\ref{radio-HI} we describe the radio
continuum and atomic hydrogen observations of the region, and we
discuss the origin of the detected sources. We present an overall
discussion of the results in Section ~\ref{discussion} before
closing with a summary in Section ~\ref{conclusions}.

\section{X-ray analysis of \object{AX~J1635.9-4719}}
\label{asca}

\subsection{Observation and Data Reduction}

The unidentified X-ray source AX J1635.9-4719 was discovered by
the {\it ASCA} telescope during a survey of the central region of
the  galactic plane, performed in the 0.7--10 keV
energy range (Sugizaki et al. \cite{sugi01}). The source is
located at $(l,\;b)=$(337.$^\circ$17, 0.$^\circ$06),
$(\alpha, \delta)_{\rm J2000.0}=(16^{\rm h}35^{\rm m}56^{\rm s}0,
-47^\circ19'54")$ (1-$\sigma$ uncertainty of $1'$). Its integrated
 flux is $\sim 1.21\times 10^{-12}$ erg cm$^{-2}$ s$^{-1}$ in the
 observed X-ray band. 

\begin{figure}[t!] 
\resizebox{\hsize}{!}{\includegraphics{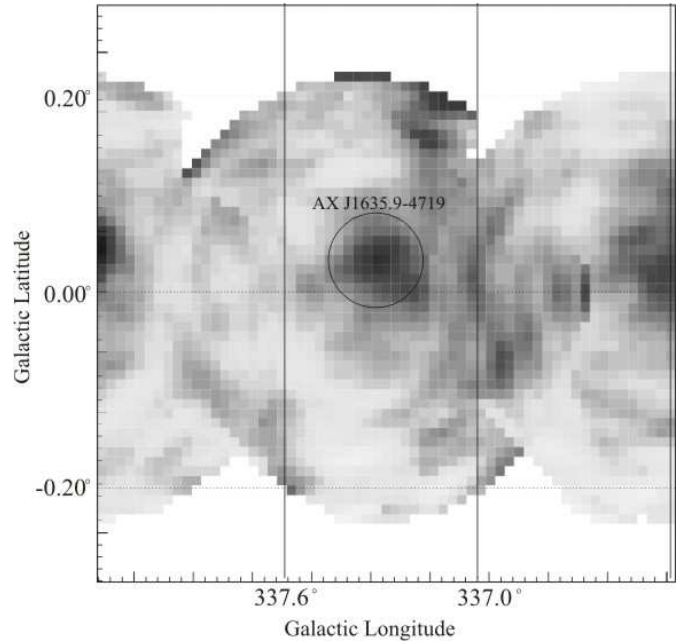}} \caption[]{{\it
ASCA} GIS mosaic image around AX J1635.9-4719 in the 2--10 keV
band. The image is smoothed with a point spread function of the
ASCA X-ray telescope with a FWHM of $3'$} \label{fig:image}
\end{figure}

\begin{figure}[t!] 
\resizebox{\hsize}{4cm}{\includegraphics{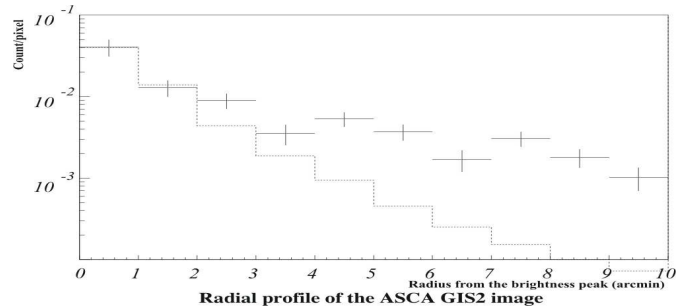}}
\caption[]{Radial profile of the X-ray image by GIS-2 around AX J1635.9-4719.
The background components of the Galactic and the extra-galactic
X-ray emissions are subtracted using the flat-field response. The intrinsic background was also subtracted following standard precedures. Dashed
line represents the radial profile expected for a point source.}
\label{fig:perfil}
\end{figure}

With the aim of studying in more detail the properties of the
X-ray source,  we have performed a more careful and thorough
imaging and spectral analysis of the ASCA data. The observation of
the source was carried out on September 7, 1997 with an exposure
of 6.1 ks. The data reduction and the extraction of the event for
the source were performed following procedures described in
Sugizaki et al. (\cite{sugi01}). Fig. ~\ref{fig:image} shows the
mosaic GIS image of AX J1635.9-4719, where the image is smoothed
with the point spread function of the ASCA X-ray telescope with a
FWHM of $3'$. The image of the source looks like a single peak
in the resolution of the instrument. To investigate the  spatial
extent of the emitting region, we extracted the radial profile of
the source in order to compare it with that of the point-spread
function of the X-ray telescope. Fig.~\ref{fig:perfil} shows the
obtained radial profile of the GIS-2 image around
\object{AX~J1635.9-4719}, where the background components of the
Galactic and the extra-galactic X-ray emissions are subtracted
using the flat-field response. Non-X-ray events were also subtracted following 
standard procedures. The central core in the inner 2
arcmin is well represented by a point source. However, an extended
component clearly exists around the source as it can be seen from
the figure.

\subsection{Light curve and spectral analysis}

\begin{figure}[t!] 
\resizebox{\hsize}{4cm}{\includegraphics{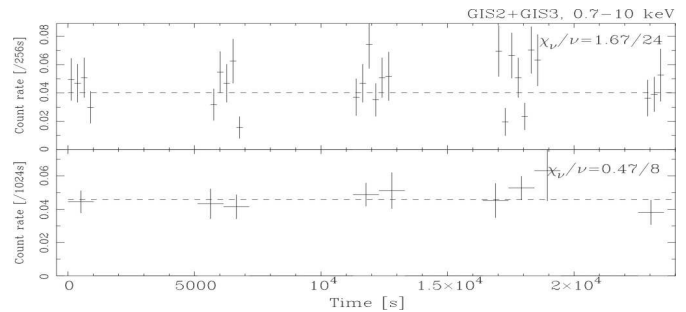}} \caption[]{X-ray
light curve of AX J1635.9-4719 during the ASCA observation on Sep.
7, 1997. There is no significant time variation. Coherent periodic
pulsation was also investigated by the FFT method. Nothing was
found in a period range of 0.5 -- 1000 s.} \label{fig:light_curve}
\end{figure}

To examine the X-ray flux variability of \object{AX~J1635.9-4719}
in more detail, we extracted a light curve and fitted it with a
constant model. There is no significant time variation. Coherent
periodic pulsation was also investigated by the FFT method: no
significant periodic signal can be seen in a period range of 0.5
-- 1000 s. Fig. \ref{fig:light_curve} shows the obtained light
curve in the whole energy band of 0.7--10 keV. The best-fit
constant model and the reduced chi-squared ($\chi^2_\nu$) of the
curve are also shown on Fig. \ref{fig:light_curve}. The source is
non-variable at a confidence above 95\%.

\begin{figure}[t!] 
\resizebox{\hsize}{4cm}{\includegraphics{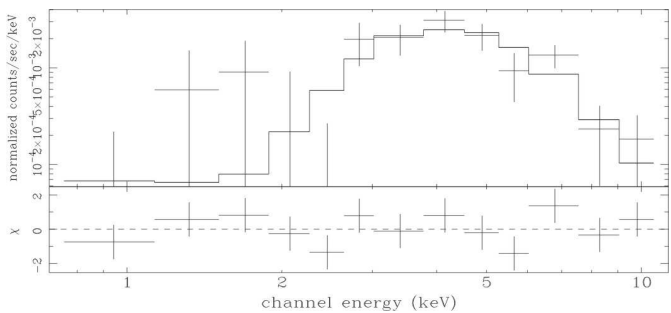}}
\caption[]{Energy spectrum observed by {\it ASCA} GIS in 0.7--10
keV.  The solid line represents the best-fit power-law model with
interstellar absorption. The photon statistics is very poor, so it
is difficult to constrain the spectral model.}
\label{fig:spectrum}
\end{figure}

We have also re-analyzed the X-ray spectrum of the source. All
spectral uncertainties represent 90\% confidence limits hereafter.
Fig. \ref{fig:spectrum} shows the obtained X-ray spectrum in the
0.7--10 keV energy range. The solid line represents the best-fit
power-law model with interstellar absorption. The photon index and
the absorption column density of the best-fit model are $\Gamma =
2.8^{+2.6}_{-1.6}$ and $N_{\rm H}=15^{+15}_{-9}\times 10^{22}$
cm$^{-2}$, respectively. The photon statistics is poor, so it is
not possible to provide a well-constrained spectral model. We also
attempted to fit thermal emission models of thin-thermal hot
plasma coded by Raymond and Smith (\cite{Raymond}) and a blackbody
model, and obtained the best-fit model statistically accepted
within the 90\% confidence level. Table \ref{tab:spectrum}
summarizes these best-fit model parameters. New observations with
better statistics are necessary to clearly separate thermal from
non-thermal emission in this source. However, it is significant
that the absorption column density is of the order of $\sim
10^{23}$ cm$^{-2}$.

\begin{table}[h]
\begin{center}
\caption{Summary of best-fit parameters of spectral fitting}
\begin{tabular}{lccc}
\hline
\hline
 & \multicolumn{3}{c}{Model}\\
Parameter & PL & Raymond$^{\rm a}$ & BB \\
\hline
$\Gamma$ or $kT$ $^{\rm b}$  & $2.8^{+2.6}_{-1.6}$ & $2.2^{+13}_{-1.1}$ & $1.4^{+0.9}_{-0.6}$ \\
$N_{\rm H}$ $^{\rm c}$    & $15^{+15}_{-9}$     & $16^{+14}_{-9}$    & $7.7^{+11}_{-5.7}$ \\
$F_{\rm 0.7-10\, keV}$ $^{\rm d}$ & 1.2                 & 1.2                & 1.2\\
\hline
$\chi^2_\nu$/d.o.f.     & 0.91/10             & 0.67/10            & 1.0/10\\
\hline
\end{tabular}
\label{tab:spectrum}
\end{center}
All errors represent the 90\% confidence limits of statistical uncertainty.\\
$^{\rm a}$ Thin-thermal plasma emission model coded by Raymond and Smith (1977) with the interstellar absorption..\\
$^{\rm b}$ Power-law photon index or temperature [keV]\\
$^{\rm c}$ Absorption column density [cm$^{-2}$] \\
$^{\rm d}$ Flux in the 0.7--10 keV band [10$^{-12}$ ergs cm$^{-2}$ s$^{-1}$] \\
\end{table}

It is worth noticing that \object{G337.2$-$0.1} lies well within
the 95\%  location contour of the unidentified $\gamma$-ray source
\object{3EG J1639-4702} (Hartman et al. \cite{hartman99}). It is
located at $(l,\; b) = (337\fdg75, -0\fdg15)$, $(\alpha,
\delta)_{\rm J2000.0} = (16^{\rm h} 39^{\rm m} 06^{\rm s},
-47\degr 02\arcmin 28\arcsec)$, and has a radius of about 0\fdg6.
Its $\gamma$-ray flux is
$(53.2\pm8.7)\times10^{-8}$~ph~cm$^{-2}$~s$^{-1}$, and presents a
steep $\gamma$-ray spectral index of $\Gamma=2.5\pm0.18$. The
EGRET source is non-variable according to both Nolan et al.
(\cite{Nolan03}) and Torres et al. (\cite{Torres01}). However,
other potential counterparts are within the error box like the
microquasar candidate AX J1639.0-4642 (Combi et al.
\cite{Combi04}) and the pulsar PSR J1637-4642 (Torres et al.
\cite{Torres03}).

\section{Radio continuum and HI observations towards \object{AX~J1635.9-4719}}
\label{radio-HI}

\object{G337.2+0.1} is a SNR candidate with an angular size of
2'$\times$3'  discovered in the MOST 843-MHz radio survey
(Whiteaok \& Green \cite{whiteoak96}). The position of the source
in galactic and equatorial coordinates is
$(l,\;b)=$(337.$^\circ$18,+0.$^\circ$06), $(\alpha, \delta)_{\rm
J2000.0}=(16^{\rm h}35^{\rm m}55^{\rm s}8, -47^\circ19'03")$
(3-$\sigma$ uncertainty of $15"$). The source has an integrated
flux density of 1.6$\pm$0.2 Jy and a mean surface brightness of
2.6$\times$10$^{-22}$ W m$^{-2}$ Hz$^{-1}$ sr$^{-1}$ at 843 MHz.
In Fig. \ref{fig:radio} we show the radio map of
\object{G337.2+0.1} at this frequency together with the 2-$\sigma$
circle of the best estimated position of \object{AX~J1635.9-4719}.

In order to find the radio source at higher frequencies we have
examined the data from the 4.85 GHz survey (Condon et al.
\cite{condon93}). Applying a filtering process to the data (see
Combi et al. {\cite{combi98} for details of the Gaussian filtering
method) we have removed the galactic diffuse emission on scales
larger than 8 arcmin and we found a weak and extended radio source
coincident with the position of \object{G337.2+0.1}. The source
has an integrated flux density of 0.39$\pm$0.02 Jy. Using the
radio flux at 843 MHz and 4.85 GHz we estimate a mean spectral
index for the source of $\sim -0.78$, confirming the non-thermal
nature of the radio source.

\begin{figure}[t!] 
\resizebox{\hsize}{!}{\includegraphics{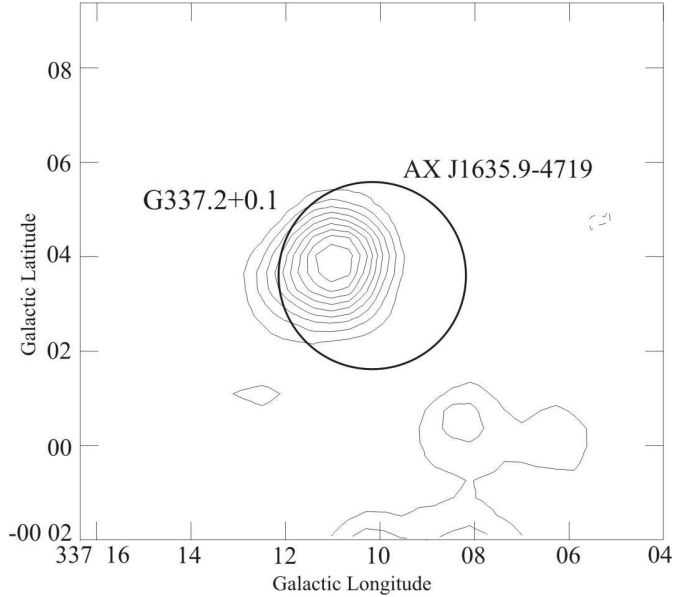}}
\caption[]{Countour image of \object{G337.2+0.1} from the MGPS
data  obtained with MOST at 843 MHz. The image size is
12'$\times$12'. Radio countours are 2, 3, 4, 5, 6, 7 and 8 times
the rms noise level of 10 mJy. The location of
\object{AX~J1635.9-4719} is also indicated.} \label{fig:radio}
\end{figure}

Since a SN explosion is expected to produce some effect in the
ISM, we  have extracted a series of HI brightness temperature maps
of the region in the velocity interval from $-100$ km s$^{-1}$ to
$+20$ km s$^{-1}$ from the Southern Galactic Plane Survey (SGPS).
These data are obtained from a combination of high-resolution
interferometric (with the Australia Telescope Compact Array) and
low-resolution (with the Parkes Radiotelescope\footnote{The ATCA
and the Parkes Radiotelescope are part of the Australia Telescope,
which is funded by the Commonwealth of Australia for operation as
a National Facility managed by CSIRO.}) observations
(McClure-Griffiths  \cite{McClure01}). The survey has angular and
velocity resolution of $\sim 2'$ and $\sim 0.82$ km s$^{-1}$,
respectively.

A series of minima can be seen in the HI channel maps, when they
are inspected at the highest velocity resolution. Only one of
these maps is exactly coincident with the radio continuum source.
The HI map integrated over the velocity range from -23 to -20 km
s$^{-1}$ is shown in Fig. \ref{fig:integra}, where we have
superposed the location of the SNR candidate.
In the HI spectrum we have found a sharp minimum at $v\sim -20$ km
s$^{-1}$. Its shape is typical of an absorption feature, hence it
can be used to set a lower limit on the distance of the background
continuum source. Using the galactic rotation studies of Russeil
(\cite{Russeil}), we get that the presumed SNR should be at least
at 13.5 kpc.

\begin{figure}[t!] 
\resizebox{\hsize}{!}{\includegraphics{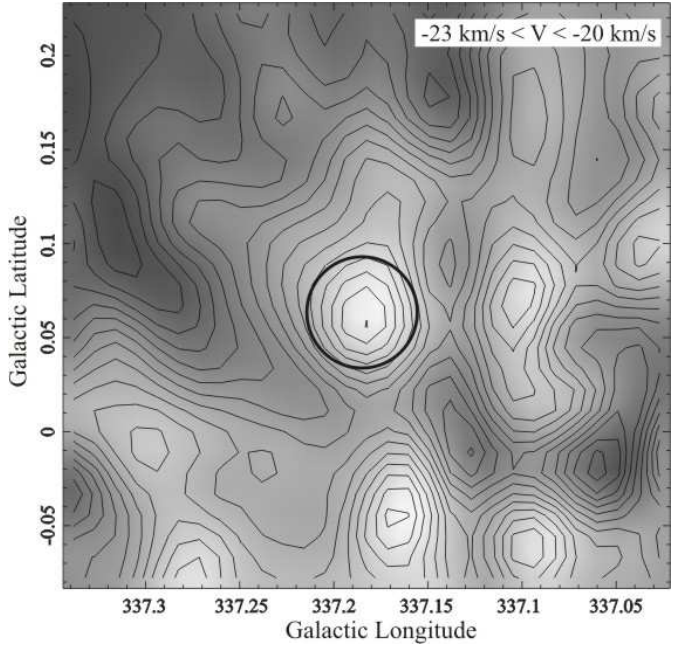}} \caption[]{HI
brightness temperature map (contour labels in K)  obtained for the
velocity range -23 to -20 km s$^{-1}$. The black circle indicate
the SNR position.} \label{fig:integra}
\end{figure}

\section{Discussion}
\label{discussion}

Several facts support the identification of the radio source \object{G337.2+0.1} with a SNR. The non-thermal
radio spectral index is typical of synchrotron emission from relativistic electrons. At X-rays the flux might
also be non-thermal but with a steeper index. This is consistent with the fact that all non-thermal X-ray SNR
spectra are substantially curved (e.g. Dyer et al. \cite{Dyer01}). The high-energy steepening of the synchrotron
spectrum can be due to the effect of losses, the presence of an exponential cut-off in the electron distribution
resulting from the failure of the acceleration mechanism at high-energies, and/or non-linear effects. In the case
of very young remnants, the particle spectrum can be limited by age (i.e. the remnant could be so young that
there was no time to accelerate electrons beyond some maximum energy). However, at the present stage a thermal
origin for the X-ray emission cannot be ruled out in this source.

The {\it ASCA} data do not allow to specify the morphology of the remnant, although seem to suggest the presence
of a central contribution. This might be emission from a pulsar or the result of the low angular resolution
and the distance to the source. We have not detected any pulsation in the data.

In addition to the broadband spectrum and the extended, non-variable emission, we have found a sharp absorption
feature  in the HI distribution toward the continuum radio source. This can be used to impose a lower bound of
$13.5$ kpc on the distance to the SNR candidate. If the actual distance is 14 kpc, the size of the remnant would
be $\sim$6 pc, similar to that of SN 1006. Assuming adiabatic expansion in a medium of density $n\sim0.5$ cm$^{-3}$,
the standard Sedov solutions (Sedov \cite{sedov59}) yield an age of $\sim$ 1500 yr and a shock front velocity of
$\sim$ 215 km s$^{-1}$.  At the same distance of 14 kpc, the X-ray luminosity of \object{AX~J1635.9-4719} is
$\sim$ 4$\times$10$^{34}$ ergs s$^{-1}$, a quite reasonable value for a young SNR.

\section{Conclusions}
\label{conclusions}

We suggest that \object{G337.2+0.1} is a SNR, being 13.5 kpc a
lower limit on its distance, and that \object{AX~J1635.9-4719} is
the X-ray counterpart of the radio source. \object{G337.2+0.1}
might have similar properties to other known galactic SNRs (e.g.
SN 1006, G266.2-1.2, G347.3-0.5 and G156.2+5.7) which exhibit a
curved broadband synchrotron spectrum. These objects have also
high X-ray luminosity, faint radio flux and they are expanding
into a low density medium (Allen et al.~\cite{allen01}; Bamba et
al. \cite{bamba01}; Koyama et al.~\cite{koya95}, ~\cite{koyama97}).

Future {\it XMM} and {\it Chandra} observations will provide a
better determination of the spectrum and morphology of this
interesting source. The electrons should also cool through inverse
Compton scattering off CMB photons (Pohl \cite{Pohl96}). This
might result in a TeV gamma-ray source that might be detected with
the {\it HESS} and CANGAROO III telescopes. Pion decays from
interactions of relativistic protons might also contribute in this
energy range.

\begin{acknowledgements}

We thank Dr. N.M. McClure-Griffiths for providing 
the HI data and useful discussions, and Dr. K. Koyama 
for useful comments. J.A.C. is a researcher of the
programme {\em Ram\'on y Cajal} funded jointly by the Spanish
Ministerio de Ciencia y Tecnolog\'{\i}a and Universidad de Ja\'en.
He was also supported during this work by CONICET (under grant PEI
6384/03). G.E.R. and P.B. were supported in part by grant BID
1201/OC-AR PICT 03-13291 (ANPCyT) and CONICET.

\end{acknowledgements}

\end{document}